 \documentclass[prd,a4paper,twocolumn]{revtex4}
 \usepackage[utf8]{inputenc} 
 \usepackage{amsmath, amssymb,mathrsfs,upgreek,graphicx}
 \usepackage{color}
 \usepackage{booktabs}
 \usepackage{multirow}
 \usepackage{bm}
\usepackage{physics} 

 \newcommand{\beq}[1]{\begin{equation}\label{#1}}
 \newcommand{\eeq}{\end{equation}}
 \newcommand{\bea}[1]{\begin{eqnarray}\label{#1}}
 \newcommand{\eea}{\end{eqnarray}}

\begin{document}

 \title{Quantum Chaology of Double Rod Pendulum}
 \author{Yu-xuan Sun}
 \email{YuXuanSun@emails.bjut.edu.cn}
 \author{Ding-fang Zeng}
 \email{dfzeng@bjut.edu.cn, ORCID: https://orcid.org/my-orcid?orcid=0000-0001-5430-0015}
 \affiliation{Institute of Theoretical Physics, Beijing University of Technology, Bejing 100124, P. R. China}
 \affiliation{Niels Bohr International Academy, Niels Bohr Institute, Blegdamsvej 17, 2100 Copenhagen, Denmark}
 \affiliation{School of Physics and Astronomy, China West Normal University, Nanchong 637002, China}
 \begin{abstract}
 The double rod pendulum is a well known classic chaotic system, so its quantum version is an ideal laboratory to test various diagnosis for quantum chaos. We quantise this system canonically and calculate its lowest $10^4$ eigenvalues and eigenstate wave functions with at least $10^{-4}$ relative precision by the spectral analysis method. With these eigenvalues and eigenstate wave functions, we calculate and examine the three popular diagnosis on quantum chaos. On the NNSD diagnosis, we find that, either the GOE feature of NNSD  is not a necessary condition for a quantum system to be chaotic at classic limit, or the double rod pendulum is not strong chaotic at the classic level. On the OTOC diagnosis, we observed that the early time exponential growth and late time constance approaching feature of OTOC is well conformed by the double rod pendulum. On the CC diagnosis, the status is similar with NNSD. Its linear growth feature at long time limit is either not a good diagnosis for a quantum system to be chaotic at classic limit or the double rod pendulum is not a strong chaotic system at classic levels.
% recomm editor, Justin David
% recomm editor, Sergey Solodukhin 
\end{abstract}
 \maketitle
 
In classical mechanics, chaos is clearly defined as the sensitive dependence of a system's evolution on initial conditions. However, due to the well known fact that quantum mechanics take wave function as the basic description for physic systems and the wave function satisfies linear Schr\"odinger equation which allows no sensitive initial-dependent evolution as time passes by \cite{Berry1981,Berry1989}, to exactly define quantum chaos is challenging question for physicists.  Since in the classic world chaos is an almost universal phenomena, especially in the many-body systems, to find quantum features of the classically chaotic system and extract universalities is a natural choice. In recent years, this question arises interests from a somewhat totally irrelevant community which lies outside the conventional complex system and complex dynamics, that is, quantum gravity professionals. Starting from the blackhole information paradoxes and stimulated partially by the Anti de-Sitter/Conformal Field Theory correspondence \cite{Maldacena:1997re}, AdS/CFT here after, Susskind \cite{Sekino2008he} and Maldacena et al \cite{MSS2016} conjecture that the black hole is the mostly chaotic object in the nature. In another word, the chaos growing speed in the nature is upper bounded and the black hole saturates this bound. 

In concrete applications, two diagnosis for ``quantum chaos'' are full-fledged, i.e., the so called NNSD (Nearest Neighbouring eigenvalue Spacing distribution) diagnosis \cite{BGS1984} and OTOC (Out of Time Ordering Correlation) diagnosis \cite{MSS2016,SSZ,SZFZ2017,FZSZ2017} . The former says that if the NNSD of a quantum system has the same feature as the Gaussian Orthogonal Ensemble (GOE) of random matrices, then that system would be chaotic at classic limit. While the latter says that in a quantum system whose classic limit is chaotic, the OTOC of two arbitrary hermitian operators from the system should exhibit short time exponential growth but long time constant approaching feature. Both these two diagnosis apply to many-body or multi-degrees of freedom systems. But in practices,  except some finitely sized lattice models, the 2-dimensional stadium billiards is almost the only calculable (numerically) model for various goals \cite{YGS,GBJMS}. So, to sharpen and deepen these diagnosis for quantum chaos, to find new calculable model or high precision numeric method is a valuable working direction. The current work will show that the double rod pendulum and spectral analysis method serves such a goal very productively.

Besides NNSD and OTOC, the CC (Circuit Complexity) is a third diagnosis for quantum chaos in developing. 
By Nielsen's geometric definition\cite{Nielsen1,Nielsen2,Nielsen3}, CC is the minimal distance from an arbitrary reference state to target state on the group manifold generated by all unitary evolution \cite{FJbook,GaussQuanInfo}. Operationally, the target state and reference state are related through $|\psi_T\rangle=e^{i H+\delta{} H t}e^{-i H t}|\psi_R\rangle$. Obviously, what CC defined this way measures is also the sensitive dependence of a quantum system's evolution on initial conditions, but with the same initial state subjected to two closely related Hamiltonian $H$ and $H+\delta{}H$. According this definition, the CC of usual harmonic oscillator is a periodical function of time, while that of the inverted harmonic oscillator grows linearly as time passes by. Considering its sensitive dependence on initial conditions, references \cite{Ali2019zcj,Qu2021ius} take the inverted harmonic oscillator a working example and believe that its linearly growing CC forms a diagnosis for quantum chaos.  However, as is well known, the inverted harmonic oscillator is exactly integrable.  Obviously, to make the linear growth of CC a firmly established diagnosis, we need to explore and test its validity/invalidity on more truly chaotic systems. The double rod pendulum provides us such an example.

The purpose of this work is to investigate the double rod pendulum's conforming/violation of the above three diagnosis for quantum chaos.  Our results will show that this system violates the first and third of them but conforms the second very well. The rest of the paper is organised as follows. Section \ref{secClassic} provides the hamiltonian formalism of the double rod pendulum and its classic evolution numerically. Section \ref{secQuantization} quantises the system canonically and introduces the spectral analysis strategy for its eigenvalue and eigenstate. Section \ref{secNNSD} studies the NNSD diagnosis of double rod pendulum as a quantum chaos. Section \ref{secOTOC} calculates several set of OTOCs in the double rod pendulum and discusses their application as the diagnosis for quantum chaos. Section \ref{secCC} generalise the definition of CC in the system and evaluates its application as a diagnosis for quantum chaos. We summary our paper in section \ref{secConcl}.

\section{Hamiltonian Formalism and Classic Dynamics}
\label{secClassic}

\begin{figure}[h]
\includegraphics[totalheight=30mm]{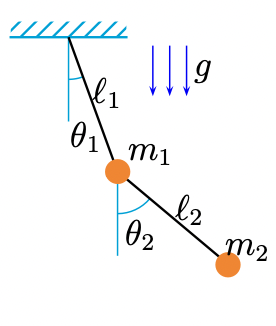}
\raisebox{-2mm}{\includegraphics[totalheight=35mm]{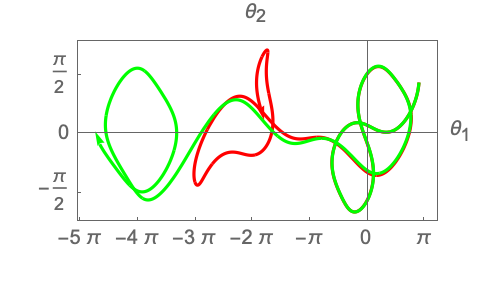}}
\\
\includegraphics[totalheight=30mm]{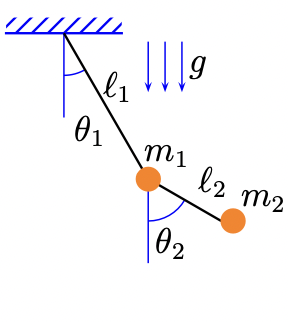}
\raisebox{-2mm}{\includegraphics[totalheight=35mm]{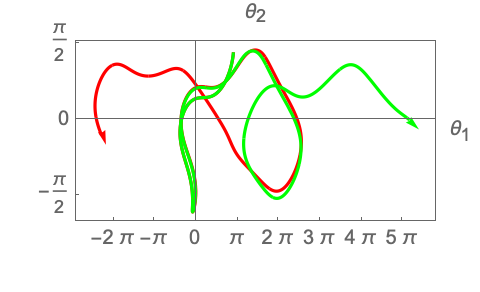}}
\\
\includegraphics[totalheight=30mm]{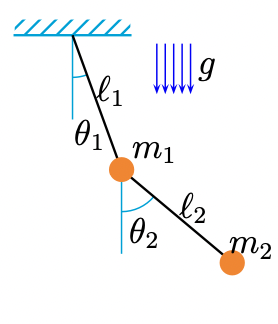}
\raisebox{-2mm}{\includegraphics[totalheight=35mm]{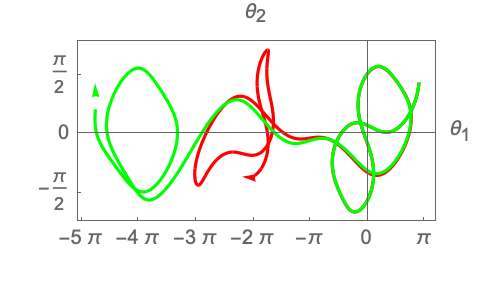}}
\caption{Classic double rod pendulums and their evolution under very close initial conditions. Masses of all three pendulums are fixed as $m_1=m_2=1$, but the upper one has $\ell_1=\ell_2=1$, $g=1$, the middle has $\ell_1=2\ell_2=\frac{4}{3}$, $g=1$, the lower has $\ell_1=\ell_2=1$, $g=10$. On arriving the displayed final configuration, three systems cost time $25,20$ and $8$ respectively. All three pairs of pendulums oscillate from very close initial conditions, $\theta_1(0)=\frac{\pi}{2}$(red) and $\frac{\pi}{2}+\epsilon(\approx0)$(green), $\theta_2(0)=\frac{\pi}{2}$, $\dot{\theta}_1(0)=0$, $\dot{\theta}_2(0)=0$}
\label{figEvolTrajectory}
\end{figure}

Referring to FIG.\ref{figEvolTrajectory}, we will characterize our double rod pendulum by the rod lengths $\ell_1$, $\ell_2$, pendulum masses $m_1$, $m_2$ and a uniform driving force field $g$ which goes directly to the floor. While angles between the two rods and driving force field direction will be denoted as $\theta_1$, $\theta_2$. With these notations, the classic action of this system reads
\bea{}
&&\hspace{-5mm} S=\int\!L(\theta_i,\dot{\theta}_i), L(\theta_i,\dot{\theta}_i)=K-V, 
\\
&&\hspace{-5mm}
K=\frac{1}{2}m_1(\ell_1\dot{\theta}_1)^2+\frac{1}{2}m_2[(\ell_1\dot{\theta}_1)^2+(\ell_2\dot{\theta}_2)^2+
\\
&&2\ell_1\ell_2\dot{\theta}_1\dot{\theta}_2\cos(\theta_1-\theta_2)]
\nonumber
\\
&&\hspace{-5mm} V=2m_1g\ell_1\sin^2\frac{\theta_1}{2}{+}2m_2g(\ell_1\sin^2\frac{\theta_1}{2}{+}\ell_2\sin^2\frac{\theta_2}{2})
\eea
In Hamiltonian formalism, this becomes
\beq{}
H(\theta_i,p_i)=\frac{p_1^2}{2I_1}+\frac{p_2^2}{2I_2}+\frac{p_1p_2}{I_{12}}+V(\theta_1,\theta_2)
,~
p_i=\frac{\partial L}{\partial\dot{\theta}_i}
\eeq
\beq{}
I_1=\big[m_1+m_2\sin^2(\theta_1-\theta_2)\big]\ell_1^2, I_2=\frac{m_2\ell_2^2 \ell_1^{-2}}{(m_1+m_2)}I_1
\eeq
\beq{}
I_{12}=m_2\ell_1\ell_2\cos (\theta_1-\theta_2)-(m_1+m_2)\ell_1\ell_2\sec (\theta_1-\theta_2)
\eeq
From this hamiltonian or Lagrangian, it's easy to derive the classic equations of motion and follow time evolutions of the system numerically. 

The double rod pendulum is a well known chaotic system. FIG.\ref{figEvolTrajectory} displays three pairs of evolution of the system with very close related initial conditions, illustrating sensitive dependence of the system's long term evolution on initial conditions. Without explicitly pointing out, all our numerical illustrations in this work will take $m_1=m_2=1$. The tunable parameters will be chosen as the relative length $\ell_1/\ell_2$ of the two arms of the pendulum and the gravitational field strength $g$. Comparing the upper and middle part of FIG.\ref{figEvolTrajectory}, we easily see that, the larger $\ell_1/\ell_2$ is, the quicker the system exhibits chaotic feature, i.e. the behaviour of the system exhibits more stronger dependence on the initial condition. Comparing the upper and lower part of FIG.\ref{figEvolTrajectory}, we see that the evolution track of the two systems are almost the same. This is caused by equivalence principle. But on arriving the same final configuration, the lower part with $g=10$ costs much shorter time. This means that the system with more larger $g$ exhibits chaotic more quickly. This is rational, because larger $g$ implies shorter characteristic time $2\pi\sqrt{\ell_\mathrm{eff}/g}$ in the system. In quantum cases, we will have an extra tunable parameter $\hbar^2$ which multiplies on all kinetic terms and controls the relative importance of kinetic energy to the potential energy in quantum models. But, the relevance of such a parameter to the system's behaviour happens always through the combination $\frac{g}{\hbar^2}$. So the effects of varying $\hbar$ is indistinguishable from those of varying $g$.

\begin{figure}[h]
\includegraphics[totalheight=40mm]{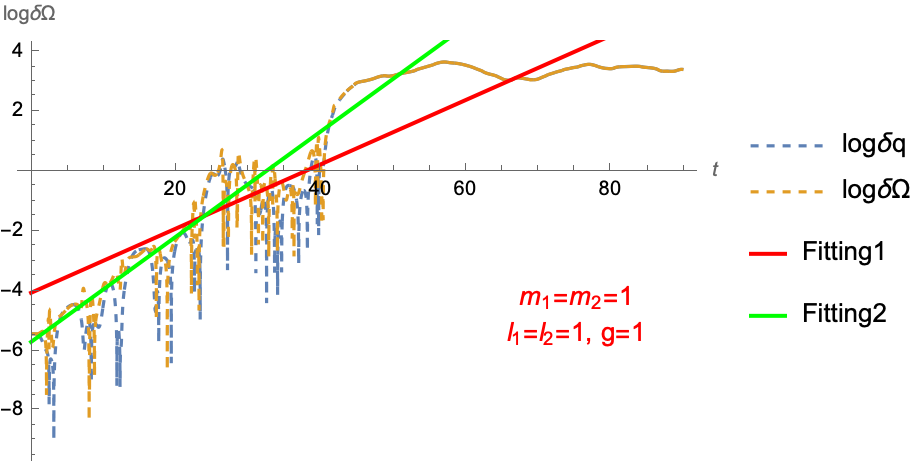}
\includegraphics[totalheight=40mm]{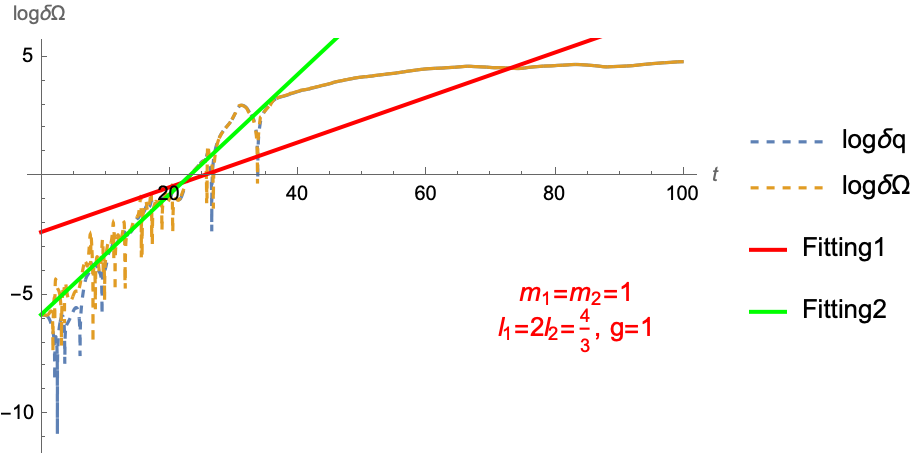}
\includegraphics[totalheight=40mm]{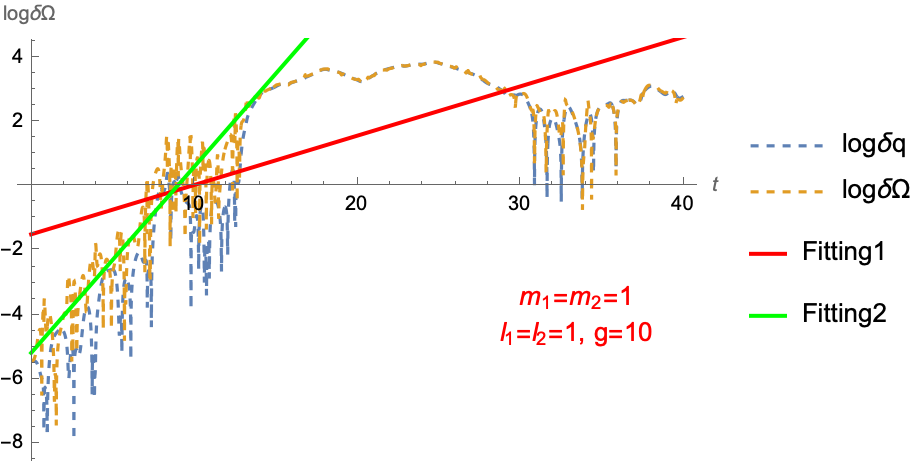}
\caption{The evolution of distance between two trajectories with slightly different initial conditions for the three double rod pendulums of FIG.\ref{figEvolTrajectory}. Red line is the fitting $\lg{\delta \Omega}$ with linear template in the whole evolution duration, while the green line seeks linear fitting only before $\delta\Omega$ grows to order $1$ value.}
\label{figLypunov}
\end{figure}

By the general definition of classic chaos \cite{FJbook,Zurek:1994wd,JRP}, the distance between two tracks of the double rod pendulum under slightly different initial conditions grows exponentially. More concretely, this implies that if $\{x(t),p(t)\}$ denote the evolution track of the system under one initial condition and $\{x(t)+\delta x(t),p(t)+\delta p(t)\}$ denote the evolution track of the system under another closely related initial, then the distance between the two tracks will grow exponentially  
\beq{}
\delta\Omega^2\equiv\delta x(t)^2+k^4\delta p(t)^2\sim\exp[\lambda_Lt]
\label{distanceInPhaseSpace}
\eeq 
where $\lambda_L$ is the so called Lyapunov exponent and $k$ is just an appropriately introduced time/length parameter aiming at dimensional balancing.  Due to the complexity of the system's dynamics, we cannot calculate the Lyapunov exponent of double rod pendulum  analytically. But we can obtain it numerically. We consider the following two examples of initial conditions
\beq{}
q_1(0)=\frac{0.99\pi}{2}, q_2(0)=0.99\pi, p_1(0)=p_2(0)=0
\label{lyinitial_conditionq1}
\eeq{}
\bea{}
q_1^{\prime}(0)=q_1(0), q_2^{\prime}(0)=q_2(0)+\epsilon
,\epsilon=10^{-6}\pi
\label{lyinitial_conditionq2}
\\
p_1^{\prime}(0)=p_2^{\prime}(0)=0
\nonumber
\eea{}
and trace the system's evolution numerically. Using the definition \eqref{distanceInPhaseSpace}, we calculate
\beq{}
 \delta\Omega^2=\bm{q}^{\prime 2}(t)+k^4 \bm{p}^{\prime 2}(t)-\bm{q}^{2}(t)- k^4 \bm{p}^{2}(t)
\eeq{}
and fit the resulting $\lg\delta\Omega$ with a linear template $a_{1}+a_{2}t$, so that the Lyapunov exponent $\lambda_L = a_2$ directly. FIG.\ref{figLypunov} tells us that the exponential growth of $\delta\Omega$ happens only in a finite time scale after the evolution begins. 

\begin{figure}[h]
\includegraphics[totalheight=27mm]{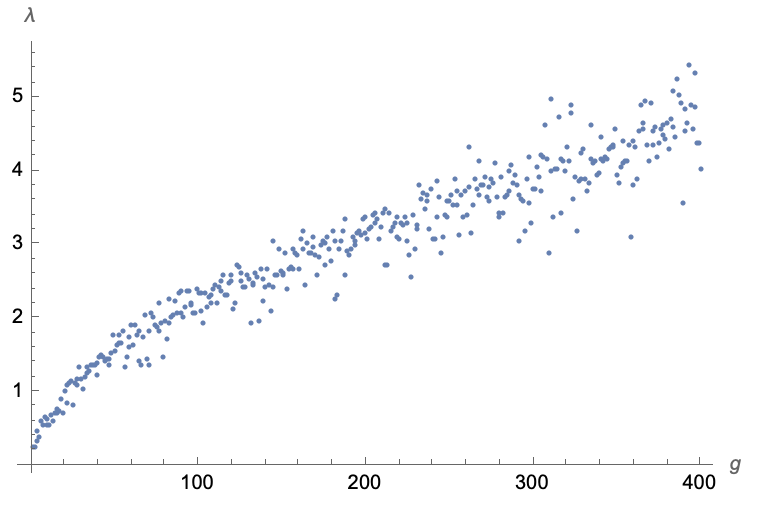}
\includegraphics[totalheight=27mm]{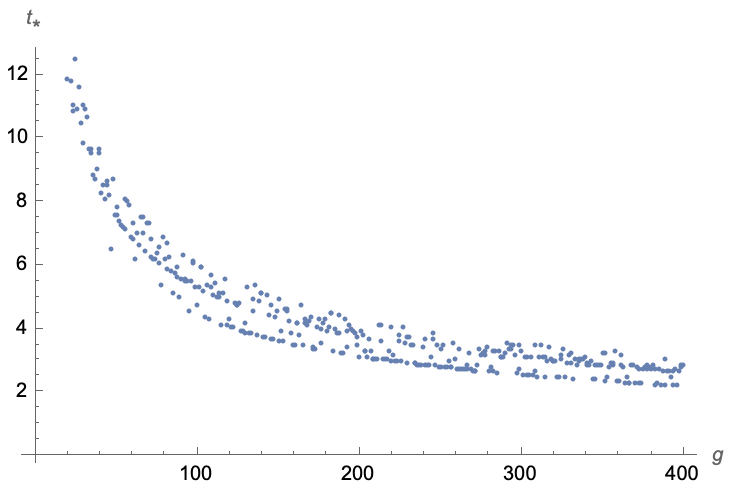}
\caption{The Lyapunov exponent $\lambda$ and scrambling time $t_*$ as the function of $g$. Model parameters are set as  $m_{1}=m_{2}=l_{1}=l_{2}=1$, $1\leqslant g\leqslant400$. In a quantised double rod pendulum, the relevance of $g$ and $\hbar$ to the system's evolution feature happens always as the combination $\frac{g}{\hbar^2}$. So the increasing of $g$ is equivalent to the decreasing of $\hbar$, and vice versa.}
\label{figLypExpAsFuncOfg}
\end{figure}

Define the scrambling time $t_*$ as the time scale on which the distance  $\delta\Omega$ between two trajectories with slightly different initial conditions grows to order $\mathcal{O}(1)$ value, FIG.\ref{figLypunov} shows that as the ratio of $\ell_1/\ell_2$ or the external field strength $g$ increases, the scrambling time $t_*$ decreases correspondingly. This is consistent with the intuitive facts indicated in FIG.\ref{figEvolTrajectory}. Physically, shorter scrambling time means stronger chaotic feature of the system. Just as we pointed out previously, when the system extends to quantum version, the effect of increasing $g$ is indistinguishable from that of decreasing $\hbar^2$. So both FIG.\ref{figEvolTrajectory} and FIG.\ref{figLypunov} tells us that as $\hbar$ increases, i.e. as $g$ decreases, the chaotic feature of the system becomes weaker and weaker, see Fig.\ref{figLypExpAsFuncOfg} for more directive illustration. This is very natural because in a fully quantised description, the system will exhibit no chaotic feature at all. This agrees with the results of \cite{Bhattacharya:1999gx,Hashimoto:2017oit} and forms challenge to the definition of quantum chaos \cite{Bhattacharya:1999gx}, and justifies at least partly the necessity of universality evaluation of NNSD, OTOC and CC diagnosis of double rod pendulum as a quantum chaos.

\section{Quantization and Numeric Strategy}
\label{secQuantization}

To make quantisation canonically, we firstly introduce a wave function  $\Psi(\theta_1,\theta_2,t)$ to describe the quantum state of the system, then replace the canonical momentum $p_1$, $p_2$ through $p_{1,2}\rightarrow i\hbar\frac{\partial}{\partial\theta_{1,2}}$, and write the hamiltonian of the system as
\beq{}
H(\theta_i,p_i){=}{-}\frac{\hbar^2}{2I_1}\!\frac{\partial^2}{\partial\theta^2_1}
{-}\frac{\hbar^2}{2I_2}\!\frac{\partial^2}{\partial\theta^2_2}
{-}\frac{\hbar^2}{I_{12}}\!\frac{\partial^2}{\partial\theta_{\!1}\!\partial\theta_{2}}
{+}V(\theta_1,\theta_2)
\eeq
Here appears our only tunable parameter $\hbar$ which controls the relative importance of the kinetic to potential energies.
The eigenvalue and eigenstate wave function will be defined through the standard timeless Schr\"odinger equation and periodic boundary conditions
\beq{}
H(\theta_1,\theta_2,\partial_{\theta_1},\partial_{\theta_2})\Psi_n(\theta_1,\theta_2)=E_n\Psi(\theta_1,\theta_2)
\label{schrodingerEquation}
\eeq 
\bea{}
\Psi(\theta_1=-\pi,\theta_2)&&\hspace{-3mm}=\Psi(\theta_1=\pi,\theta_2)
\label{periodicBoundaryConditionA}\\
\Psi(\theta_1,\theta_2=-\pi)&&\hspace{-3mm}=\Psi(\theta_1,\theta_2=\pi)
\label{periodicBoundaryConditionB}\\
\partial_{\theta_1}\!\Psi(\theta_1=-\pi,\theta_2)&&\hspace{-3mm}=\partial_{\theta_1}\!\Psi(\theta_1=\pi,\theta_2)
\label{periodicBoundaryConditionC}
\\
\partial_{\theta_2}\!\Psi(\theta_1,\theta_2=-\pi)&&\hspace{-3mm}=\partial_{\theta_2}\!\Psi(\theta_1,\theta_2=\pi)
\label{periodicBoundaryConditionD}
\eea
Our numeric strategy to solve this system is as follows. Firstly, we uniformly discretize the $\theta_1,\theta_2$ coordinates into two $N_1, N_2$ dimensional vectors $\vec{\theta}_1\equiv\theta_1^i$, $\vec{\theta}_2\equiv\theta_2^j$ and represent the first and second order differential operators ${\partial/\partial\theta_1}$, $\partial^2/\partial\theta_1^2$(${\partial/\partial\theta_2}$, ${\partial^2/\partial\theta_2^2}$ similarly) as the skew- and normal-symmetric $N\times N$ Toeplitz matrix such as
\bea{}
&&\hspace{-1mm}D_1=\mathrm{Toeplitz}^\mathrm{skew}[\{\{0\},\frac{\!(-1)^{1:N_1}\!\!}{2}\!\cot\frac{1{:}N_1\pi}{N_1{+}1}\}]
\label{toeplitzFirstOrder}\\
&&\hspace{-5mm}\mathbb{D}_1=\mathrm{Toeplitz}^\mathrm{normal}[\{\{-\frac{N_1^2}{12}-\frac{1}{6}\},\frac{\!-(-1)^{1:N_1}\!\!}{2\sin^2(\frac{1:N_1\pi}{N_1+1})}\}]
\label{toeplitzSecondOrder}
\eea
Secondly, we tensor product $\vec{\theta}_1$, $\vec{\theta}_2$ to get the extended coordinate vector
$\theta^k\equiv\theta_1^i\otimes\theta_2^j$ ($k{=}i{\cdot}N_1{+}j$) and represent the wave function $\Psi(\theta_1,\theta_2)$ with its values on this extended vector correspondingly. This means that $\Psi(\theta_1,\theta_2)$ also becomes vector of dimension $N_1{\cdot}N_2$. Correspondingly the Schrodinger equation 
\eqref{schrodingerEquation} will be written into an $(N_1{\cdot}N_2)\times(N_1{\cdot}N_2)$ linear algebra equation array
\bea{}
\big[{-}\frac{\hbar^2}{2I_1}\!\mathbb{D}_1\otimes\mathbb{I}_{N_2{\times}N_2}
{-}\frac{\hbar^2}{2I_2}\!\mathbb{I}_{N_1{\times}N_1}\otimes\mathbb{D}_2
{-}\frac{\hbar^2}{I_{12}}\!D_1{\otimes}D_2
\label{SchrodingerEqMatrixForm}
\\
{+}V(\theta_1,\theta_2)\big]_{k\ell}\Psi_\ell=E_k\Psi_k
\nonumber
\eea
The Toeplitz definition of the differential operators \eqref{toeplitzFirstOrder}-\eqref{toeplitzSecondOrder} assures that the solutions to the equation array \eqref{SchrodingerEqMatrixForm} will satisfy boundary conditions \eqref{periodicBoundaryConditionA}-\eqref{periodicBoundaryConditionD} naturally. The question here is that, all coefficients here $I_1$, $I_2$, $I_{12}$ and the potential $V$ are functions of $\theta\equiv\vec{\theta}_1\otimes\vec{\theta}_2$ thus should be considered diagonal matrices. When $I_1$, $I_2$ and $I_{12}$ times on the corresponding differential matrices, they will break symmetries of the Hamiltonian under subscript exchanging, thus breaking their hermiticity. This is the appearance of the universal normal ordering problems when quantizing interacting systems with more than one degrees of freedom. To avoid this fatal events, we choose symmetrize all the product of the $I$-coefficients and their differential operators in eq\eqref{SchrodingerEqMatrixForm} to get the proper eigenvalue and eigenvector.

\begin{figure}[h]
\includegraphics[totalheight=23mm]{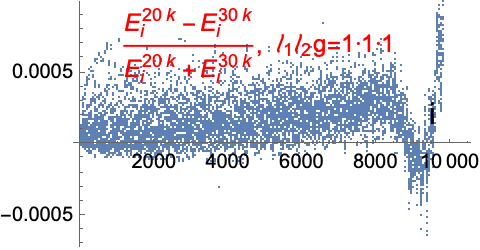}
\includegraphics[totalheight=23mm]{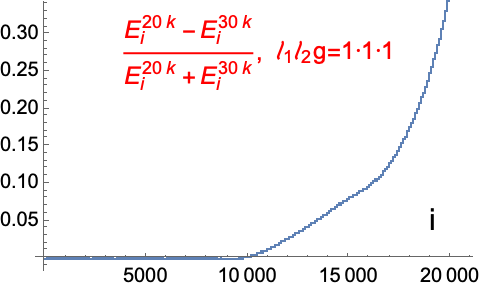}
\\
\includegraphics[totalheight=23mm]{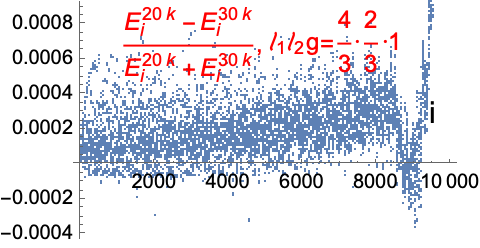}
\includegraphics[totalheight=23mm]{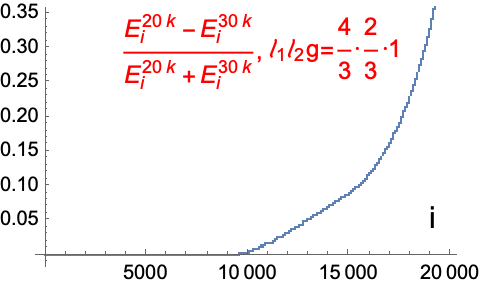}
\\
\includegraphics[totalheight=23mm]{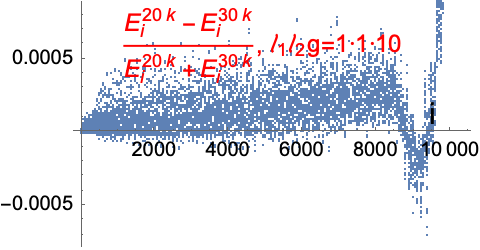}
\includegraphics[totalheight=23mm]{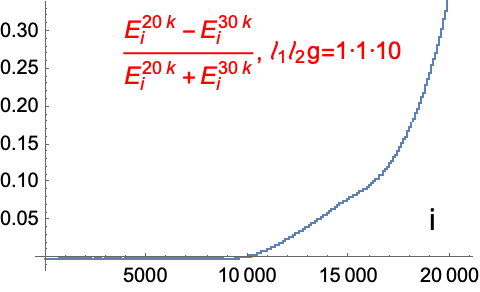}
\caption{Eigenvalue error's estimation through ratios of difference over summation of eigenvalues following from 20$k$($2\times10^4$) and 30$k$($3\times10^4$) components implementation of equation \eqref{SchrodingerEqMatrixForm}.}
\label{figEiErrorEstimation}
\end{figure}

With the power of mainstream personal computer or mac-books on the current market, we can construct matrix of the hamiltonian in equation \eqref{SchrodingerEqMatrixForm} as large as $30k\times30k$ and solve its $30k$ eigenvalues in half an hour. However, only the lowest half of these eigenvalues is reliable. The higher half usually contain errors of $\mathcal{O}(0.1)$ due to too sparse sampling of the wave function on the coordinate space $\vec{\theta}\equiv\vec{\theta}_1\otimes\vec{\theta}_2$. Use the ratio of difference over summation of eigenvalues computed from 20k and 30k components implementation of eq\eqref{SchrodingerEqMatrixForm}, we plot in FIG.\ref{figEiErrorEstimation} error estimations of our calculation directly. From the figure it's easy to see that, for all the three parameter sets chosen in FIG.\ref{figEvolTrajectory}, our results for at least the lowest 10000 eigenvalues in each cases are reliable, which means that the relative error is of order $10^{-4}$. However, for our statistic analysis of the system's eigenvalue distribution, this number is enough. If we want to calculate more eigenvalues of the system, the memory and computation time for implementing eq\eqref{SchrodingerEqMatrixForm} will grow cubically as the expected number of eigenvalues increase. In the high energy limit, the eigenvalue spectrum is roughly linear in the energy level n and exhibits no degeneracy. The spectrum can be well fitted with linear template $E_n\sim 0.13293n+2.78853 (l_1/l_2=1, m_1/m_2=1, g=1)$ which agrees with ref.\cite{DoublePdlm} very well.

\begin{figure}[h]
\includegraphics[totalheight=25mm]{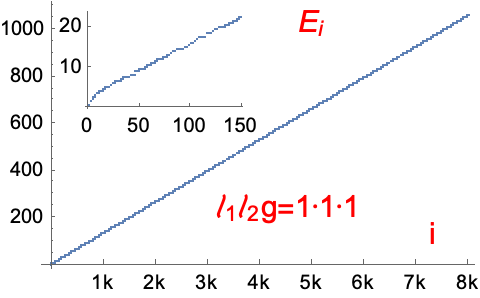}
\includegraphics[totalheight=25mm]{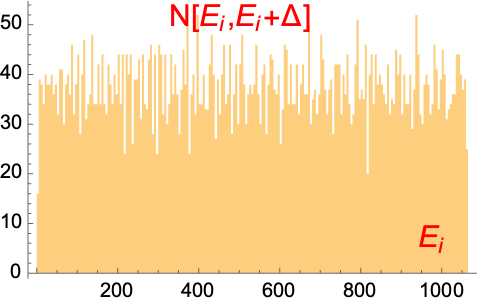}
\\
\includegraphics[totalheight=25mm]{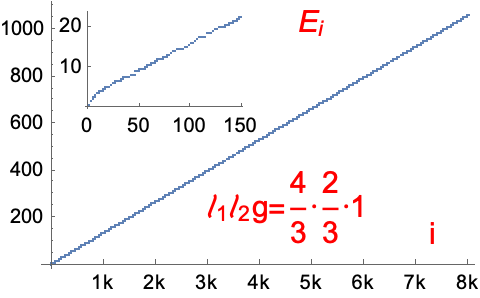}
\includegraphics[totalheight=25mm]{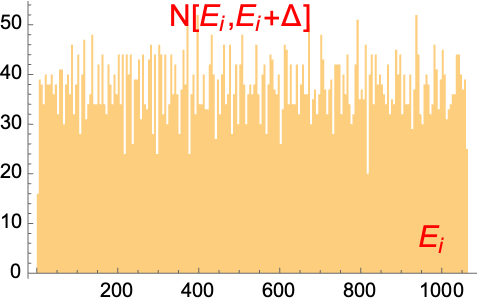}
\\
\includegraphics[totalheight=25mm]{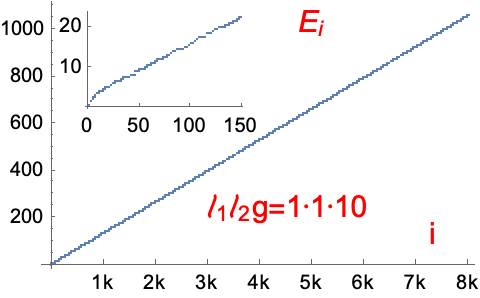}
\includegraphics[totalheight=25mm]{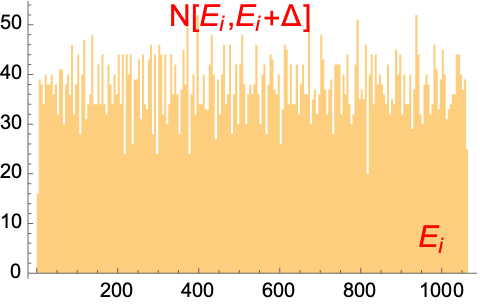}
\caption{List and distribution of the eigenvalue of quantum double rod pendulum. }
\label{figEiList}
\end{figure}

\begin{figure}[h]
\includegraphics[totalheight=25mm]{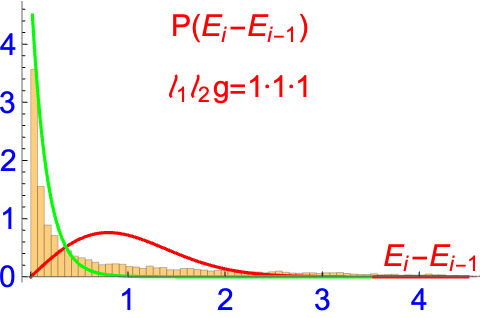}
\includegraphics[totalheight=25mm]{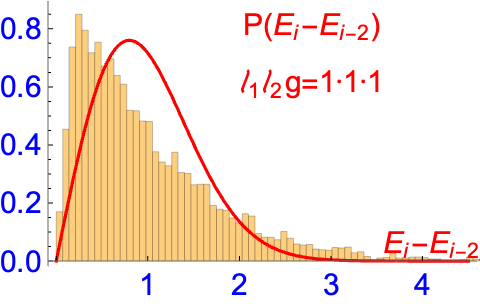}
\\
\includegraphics[totalheight=25mm]{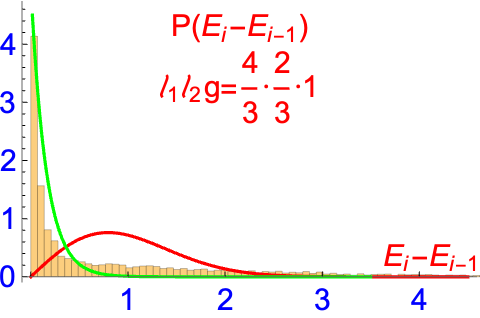}
\includegraphics[totalheight=25mm]{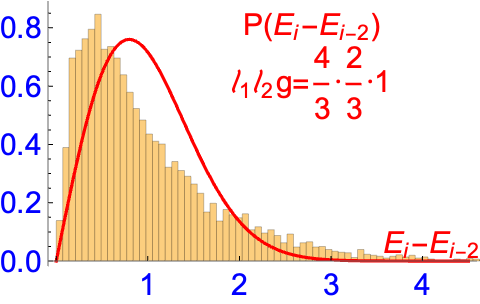}
\\
\includegraphics[totalheight=25mm]{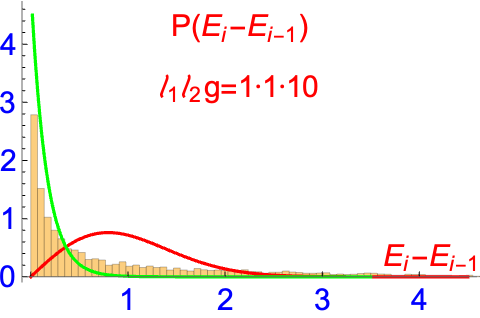}
\includegraphics[totalheight=25mm]{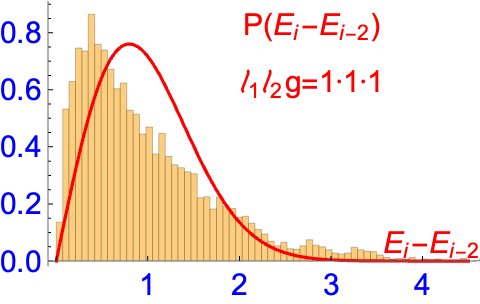}
\caption{The left is the distribution of the NNSD, i.e. Nearest Neibouring eigenvalue Spacing Distribution, while the right is the Next-to-Nearest Neibouring eigenvalue Spacing Distribution of the quantum double rod pendulum  The red line is the expected GOE distribution, the green line is the poisson distribution. 
}
\label{figNNSD}
\end{figure}

\section{NNSD Diagnosis}
\label{secNNSD}

We provide in FIG.\ref{figEiList} and \ref{figNNSD} distributions of the eigenvalue and its nearest neighboring spacing, NNSD for simplicity. The former is a global feature of the system's eigenstate spectrum, i.e. its eigenstate number density as a function of the excitation energy. While the latter is a local property of the eigenstate spectrum. It is widely believed that the NNSD of quantum chaotic systems has the same feature as that of the Gaussian Orthogonal Matrix Ensembles, while that of the integrable systems has the Poisson feature \cite{BGS1984,CHS,JMG,SWM,RUH}. 
This so called BGS (Bohigas, Giannoni and Schmit ) conjecture is usually considered a standard diagnosis for quantum chaos. Typical examples supporting this conjecture include stadium billiards and experimentally obtained nuclear resonance energy-level data. In FIG.\ref{figNNSD}, we write,\bea{}
&&\hspace{-5mm}p_{\scriptscriptstyle\mathrm{GOE}}(x)=\frac{\pi\,x}{2}\,\exp(-\frac{\pi x^2}{4})
\\
&&\hspace{-5mm}p_{\scriptscriptstyle\mathrm{Poisson}}=5\exp(-2\pi\,x)
\eea
The numerics here in the Poisson distribution is determined by hands so that the resulting distribution fit the numerical results well, while those in the GOE distribution is determined by normalisation with $x$ ranges in the whole positive semi-axis.

From FIG.\ref{figNNSD}, we easily see that the NNSD of the double rod pendulum has Poisson distribution instead of GOE. This seems to form a counter example to the BGS conjecture \cite{BGS1984}, which says that ``Spectra of time reversal-invariant systems whose classical analogs are K systems show the same fluctuation properties as predicted by GOE". The double rod pendulum is obviously time reversal-invariant. So the fact that its NNSD exhibits no GOE feature implies that, either it is not a K system, i.e. strong chaotic system or it forms a counter example to the BGS conjecture. Ref.\cite{NNSDPoison} argues that, the NNSD of integrable systems with more than one degrees of freedom should be poisson like. The fact that the double rod pendulum has poisson like NNSD but chaotic at classic limits implies that this argument may provide only necessary but not sufficient judgement for the integrability of a system. 

Although the NNSD of double rod pendulum exhibits no GOE feature, we notice that NNNSD $E_i-E_{i-2}$, i.e. the Next Nearest Neiboring eigenvalue's Spacing Distribution exhibits GOE feature, see the right panel of FIG.\ref{figNNSD}.  As is known, the eigenvalue spacing's distribution of Gaussian Orthogonal Matrix Ensembles exhibits Gaussian feature whatever number of level-spacing we consider. So it may seems a reasonable idea to extend the BGS conjecture to introduce the higher separation of eigenvalue spacing's distribution as a diagnosis for the chaotic feature of a quantum system at classic limits. Ref \cite{Unfolding2} claims that the unfolding of eigenvalue's distribuiton is not trivial and can change the feature of resulting distribution of the system remarkably \cite{Unfolding1,Riser:2020sdn}. But we checked this point and find that it does not change our conclusion qualitatively.

In a word, our results in this section implies that either the double rod pendulum is not a strong chaotic system or it forms a counter example to the BGS conjecture and the GOE feature of NNSD is not a necessary diagnosis for a quantum system's chaotic feature at classic limit. 
 
\section{OTOC Diagnosis}
\label{secOTOC}

In classic mechanics, the difference between two evolution tracks of a chaotic system with slightly different initial conditions grows exponentially $\frac {\delta x(t)}{\delta x(0)}\sim e^{\lambda_{L}t}$ as time passes by, where $\lambda_L$ is the Lyapunov exponent \cite{FJbook,Zurek:1994wd,JRP}. In quantum mechanics, $\frac{\delta{}x(t)}{\delta{}x(0)}\rightarrow{}i\hbar^{-1}[x(t),p(0)]$, so the classic saying $\frac {\delta x(t)}{\delta x(0)}\sim e^{\lambda_{L}t}$ becomes $\langle[x(t),p(0)]^2\rangle\sim\hbar^2e^{2\lambda_Lt}$. For more general systems, the square of two arbitrary hermitian operator's unequal time commutator $C(t)=\langle[W(t),V(0)]^2\rangle$ becomes the natural choice of diagnostic quantity for quantum chaos \cite{MSS2016,1969JETP,Maldacena:2016hyu,KitaevKITP}.  Historically, the OTOC function $F(t)$ refers to exactly the out of time ordering correlation of two operators like $W$ and $V$
\beq{}
F(t)\equiv\langle{}W(t)V(0)W(t)V(0)\rangle_\beta
\eeq
where $\langle\cdots\rangle_\beta$ means averaging over thermal ensemble of the system at temperature $\beta^{-1}\equiv k_{\scriptscriptstyle\!B}T$. This definition of OTOC is related with the unequal time commutator $C(t)$ as follows
\bea{}
&&\hspace{-5mm}C(t)\equiv-\langle[W(t),V(0)]^2\rangle_\beta=-\langle{}W_tV_0W_tV_0+
\\
&&\hspace{1mm}V_0W_tV_tW_0-W_tV_0V_0W_t-V_0W_tW_tV_0{}\rangle_\beta
\nonumber
\\
&&\hspace{-5mm}=\!-[2F(t)-2D(t)]
\eea   
Kiataev proposes the first OTOC diagnosis for quantum chaos in reference \cite{1969JETP,Kitaev2014}, which says that the short term growth of a quantum system's OTOC function $F(t)$ will be exponential if its classic limit is chaotic. While references \cite{Almheiri:2013hfa,MSS2016} argue that the long term feature of $C(t) \sim \left\langle VV \right\rangle \left\langle WW \right\rangle\rightarrow$constant or the long term feature of $F(t)\sim0$ may also be used as a diagnosis for a quantum system's chaotic feature at classic limit.

\begin{figure}[h]
\includegraphics[totalheight=25mm]{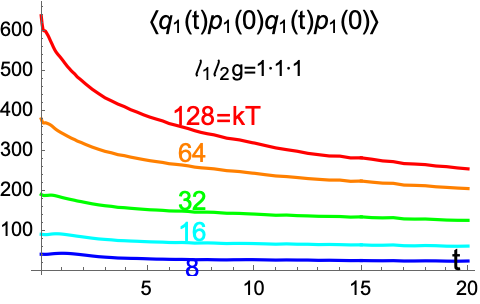}
\includegraphics[totalheight=25mm]{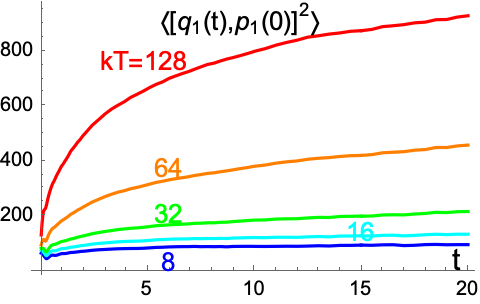}
\includegraphics[totalheight=25mm]{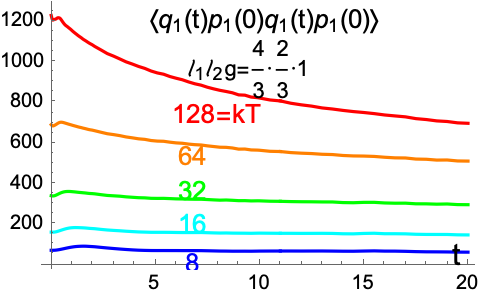}
\includegraphics[totalheight=25mm]{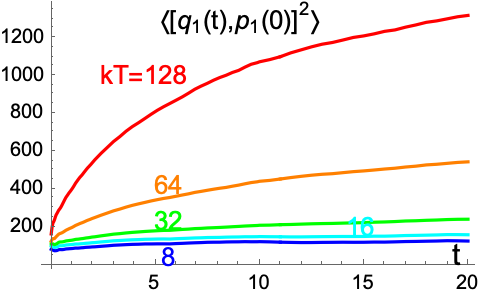}
\includegraphics[totalheight=25mm]{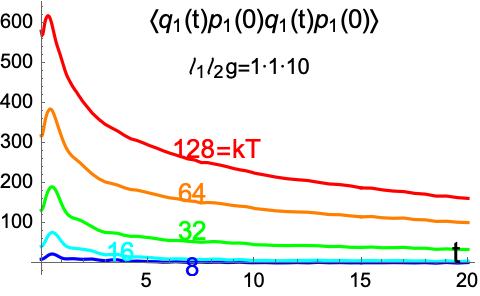}
\includegraphics[totalheight=25mm]{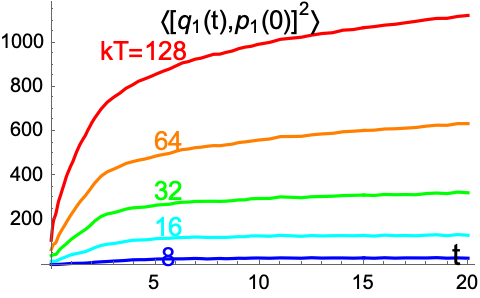}
\caption{The long term behaviour of OTOC $F$ and $C$ function of the three double rod pendulums of FIG.\ref{figEvolTrajectory}}
\label{figOTOCLongTMfeature}
\end{figure}

For the quantum double rod pendulum, through our eigenstate wave function obtained numerically in section \ref{secQuantization}, we can easily calculate the operator matrix elements such as $\theta^1_{nk}=\langle\Psi_n|\theta_1|\Psi_k\rangle$ and $p^1_{k\ell}=\langle\Psi_n|p_1|\Psi_k\rangle$ et al. With these matrix elements, the OTOC of $\theta^i$ and $p^j$ can be calculated routinely as,
\bea{}
&&\hspace{-5mm}F(t)=\langle \theta_1(t)p_1(0)\theta_1(t)p_1(0)\rangle
\label{OTOCdefinition}
\\
&&\hspace{-5mm}=\!\sum_{n=0}^\infty\!e^{\!-\beta E_n}\!\langle\Psi_n|e^{iHt}\theta_1e^{\!-iHt}p_1e^{iHt}\theta_1e^{\!-iHt}p_1|\Psi_n\rangle
\\
&&\hspace{-3mm}=\mathrm{tr}\,e^{-\beta E_n+iE_nt}\theta^1_{nk}e^{\!-iE_kt}p^1_{k\ell}e^{iE_{\ell}t}\theta^1_{\ell{}m}e^{\!-iE_mt}p^1_{mn}
\eea
\bea{}
\label{OTOCnormalization}
C(t)=2\mathrm{tr}\,e^{-\beta E_n}\theta^1_{nk}e^{\!-iE_kt}p^{1{\cdot}2}_{k\ell}e^{iE_{\ell}t}\theta^1_{\ell{}n}-2F(t)
\eea
where $\theta^1_{nk}=\langle\Psi_n|\theta_1|\Psi_k\rangle$ and $p^1_{k\ell}=\langle\Psi_n|p_1|\Psi_k\rangle$ et al are calculated through the coordinate representation of $\theta_1$, $p_1$ and the eigenstate function. For $\theta_2$ and $p_2$ we have expressions completely similar with \eqref{OTOCdefinition}-\eqref{OTOCnormalization}. Our results are displayed in FIG.\ref{figOTOCLongTMfeature} and \ref{figOTOCShortTMfeature} explicitly.
\newcommand{\tableQLyapnovFittingA}{\begin{tabular}{cccc}
$a$ & $b$ & $\lambda^q_L$ & $\frac{2\pi}{\beta}$
\\
\hline
46.43 & 13.06 & 2.50 & $2^8\pi$
\\
42.64 & 12.06 & 2.57 & $2^7\pi$
\\
34.72 & 9.92 & 2.74 & $2^6\pi$
\\
20.88 & 6.24 & 3.15 & $2^5\pi$
\\
7.46 & 2.66 & 3.83 & $2^4\pi$
\end{tabular}}
\newcommand{\tableQLyapnovFittingB}{
\begin{tabular}{cccccc}
$a$ & $b$ & $\lambda^q_L$ & $\frac{2\pi}{\beta}$
\\
\hline
53.59 & 15.38 & 8.10 & $2^8\pi$
\\
42.88 & 13.32 & 8.20 & $2^7\pi$
\\
23.26 & 8.72 & 8.41 & $2^6\pi$
\\
6.13 & 3.26 & 8.70 & $2^5\pi$
\\
1.28 & 0.95 & 8.40 & $2^4\pi$
\end{tabular}}
\newcommand{\tableQLyapnovFittingC}{
\begin{tabular}{cccc}
$a$ & $b$ & $\lambda^q_L$ & $\frac{2\pi}{\beta}$
\\
\hline
42.59 & 24.15 & 11.61 & $2^8\pi$
\\
26.23 & 16.72 & 11.72 & $2^7\pi$
\\
7.88 & 7.11 & 11.48 & $2^6\pi$
\\
1.28 & 2.28 & 10.13 & $2^5\pi$
\\
0.21 & 0.82 & 8.74 & $2^4\pi$
\end{tabular}}
\begin{figure}[h]
\includegraphics[totalheight=29mm]{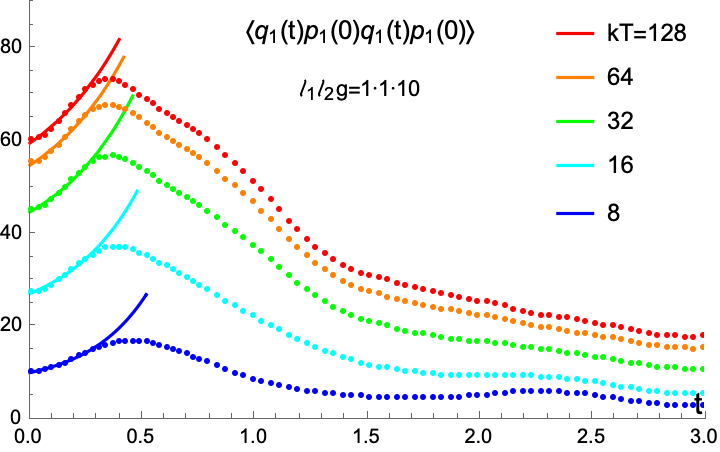}
\raisebox{16mm}{\tableQLyapnovFittingA}
\\
\includegraphics[totalheight=29mm]{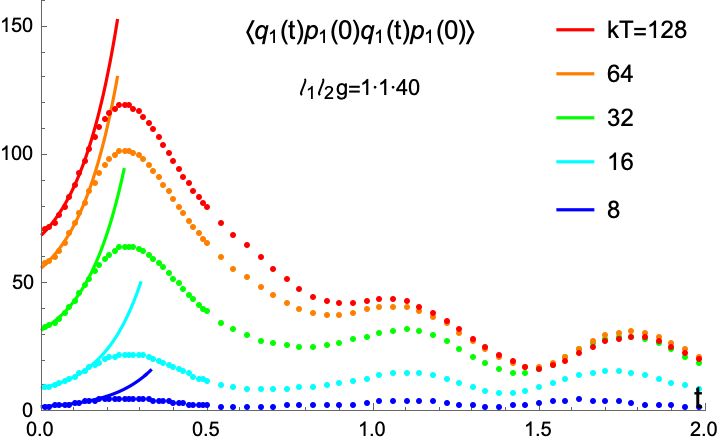}
\raisebox{16mm}{\tableQLyapnovFittingB}
\\
\includegraphics[totalheight=29mm]{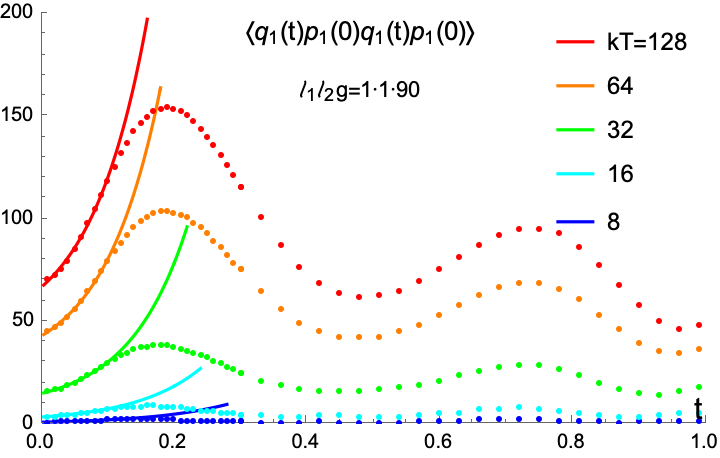}
\raisebox{16mm}{\tableQLyapnovFittingC}
\caption{The short term behaviour of the OTOC $F$-function and its exponential fitting with template $a+be^{\lambda^q_Lt}$ in three double rod pendulums. All model parameters are displayed in the figure.}
\label{figOTOCShortTMfeature}
\end{figure}

FIG.\ref{figOTOCLongTMfeature}  shows that in the long time limit, the OTOC $F$-function decays to zero asymptotically while the $C$-function grows to constant. This forms perfect support for the argument of ref.\cite{Almheiri:2013hfa,MSS2016,PRLOTOC}. As long as the short term behaviour is concerned, our results indicate that the exponential growing feature of OTOC $F$-function is observable only in the large enough $g\rightarrow\infty$ or small enough $\hbar\rightarrow0$ systems, see FIG.\ref{figOTOCShortTMfeature} for illustration. This again confirms the well known fact that chaos happens only in classic systems and all isolated quantum systems allow no chaotic evolution. Besides this, our results also provide support for Kiataev's OTOC diagnosis very well but form contrasts to the result of ref.\cite{Hashimoto:2017oit} which sees no short term exponential growth of $F$ in the stadium billiard model. On the fitting of FIG.\ref{figOTOCShortTMfeature}, we need to notice two points. The first is $\lambda^q_L$ denotes the Lyapunov exponent of the quantum system. It may be linearly related with but not identically equal to the classic $\lambda_L$ displayed in FIG.\ref{figLypExpAsFuncOfg}. The second is that we fixed the range of numeric fitting on each diagram to their first ten scattered points artificially. This may cause remarkable error for the resulting $\lambda^q_L$. But the qualitative trend that $\lambda^q_L$ grows proportionally with $g$ is reliable.

In ref.\cite{MSS2016}, Maldacena, Shenker and Stanford conjecture that the growth of $C(t)$ is upper bounded $\frac{dC}{dt}\leqslant \frac{2\pi}{\beta} C$ if the following two conditions hold: (i) the system considered contains a large gap between $t_d$ and $t_*$, where $t_d$ is the exponential decay time on which the general two-point correlation function decays significantly and $t_*$ is the scrambling time on which $C(t)$ grows near its upper bound; (ii) on time scales $t_d \ll t \ll t_*$, OTOC can be roughly factorized. Combining with Kitaev's argument \cite{Kitaev2014} that a chaotic system has $C(t)\sim{}e^{\lambda^q_L t}$, MSS conjecture tells us that $\lambda^q_L \leqslant \frac{2\pi k_B T}{\hbar}$.  This bound was originally suggested in the study of black hole information paradoxes \cite{Shenker:2013pqa,Shenker:2013yza,Kitaev2014,Polchinski:2015cea,Jackson:2014nla}, with the exactly solved Sachdev-Ye-Kiataev(SYK) model \cite{Sachdev:1992fk,KitaevKITP,Kitaev:2017awl} saturate it exactly. Our results in FIG.\ref{figOTOCShortTMfeature} indicate that the double rod pendulum conforms this bound very well with the corresponding $\lambda^q_L$ lies far below the saturation limit. However, as $g$ becomes larger and larger, the error of numeric calculation becomes more and more difficult to control. So our results do not definitely tell us if the upper bound would be saturated at large $g$. Remember that, large $g$ implies small $h$ thus classic limit. Just as FIG.\ref{figLypExpAsFuncOfg} tells us, in such limit the corresponding $\lambda^q_L$ grows indeed but numeric work cannot answer us definitely if it has an upper bound or not. We accept that this is a shortcoming of pure numeric works.

\section{CC Diagnosis}
\label{secCC}

Ref \cite{Ali2019zcj} proposes the idea of using circuit complex as another alternative diagnosis for quantum chaos, which defines the circuit complex as follows
\bea{}
\mathcal{C} & =&\frac{1}{2\sqrt{2}}\sqrt{\text{Tr}\left[\left(\log\Delta\right)^{2}\right]}
\label{aliCCdefinition}
\\
\Delta & =& G_{\text T}G_{\text R}^{-1}
\label{aliDltMatdefinition}
\eea
where $G_\text R, G_\text T$ denote the covariant matrix of the reference and target states respectively. For simplicity, ref\cite{Ali2019zcj} chooses a gaussian state $\psi$ as reference and an unitary gaussian evolution operator $U$ to define the covariant matrix: 
\bea{}
G_\text{R}=\bra{\psi}\xi_i\xi_j\ket{\psi}, \quad G_\text{T}=\bra{\psi^\prime}\xi_i\xi_j\ket{\psi^\prime}
\label{aliGRGTdefinition}
\eea
where $\xi_i\equiv\{x(t)/k,kp(t)\}$ with $k$ is a dimensional balancing constant and $\ket{\psi^\prime}=U\ket{\psi}$, $U=e^{iH't}e^{-iHt}$. In the simple harmonic oscillator and reverted harmonic oscillator considered by ref.\cite{Ali2019zcj}, $H'=H[\omega\rightarrow(1+\epsilon)\omega]$. This choice of $U$ assures that a gaussian initial state will evolve to a gaussian final one. In the simple harmonic oscillator and inversed harmonic oscillator, this definition of circuit complex can be proved equivalent with the geometric definition of \cite{FJbook,Ali2019zcj,Nielsen1,Nielsen2,Nielsen3,Chapman:2018hou,Bhattacharyya:2018bbv}.

The most remarkable difference between the double rod pendulum and the simple harmonic or inverted harmonic oscillator is the former's nonlinearity. This feature will cause difficulties to keep a reference gaussian state be gaussian as evolution happens. However, as a simple but natural generalisation, we can define the reference state as the ground state of $H$ which is approximately gaussian and the target state as follows
\bea{}
|\psi'\rangle=U|\psi\rangle, U=e^{iH't}e^{-iHt}\rule{8mm}{0pt}
\label{evolUDRP}
\\
H'=H[\ell_1\rightarrow(1+\epsilon)\ell_1,\ell_2\rightarrow(1-\epsilon)\ell_2]
\label{HprimeDRP}
\eea
Of course the phase space correspondingly generalises to four dimensional
\beq{}
\xi_i=\{\theta_1(t),\theta_2(t),kp_1(t),kp_2(t)\}
\label{xiDRP}
\eeq
In numerical works, we take $\epsilon=10^{-6}$, while the dimensional balancing parameter $k$ will be chosen as the quasi-periodic time of our system $2\pi(\ell_{eff}/g)^\frac{1}{2}$ \cite{DoublePdlm}. With these preparations, we can calculate the circuit complexity of the system following eqs\eqref{aliGRGTdefinition},\eqref{aliDltMatdefinition} and \eqref{aliCCdefinition}.

\begin{figure}[h]
\includegraphics[totalheight=25mm]{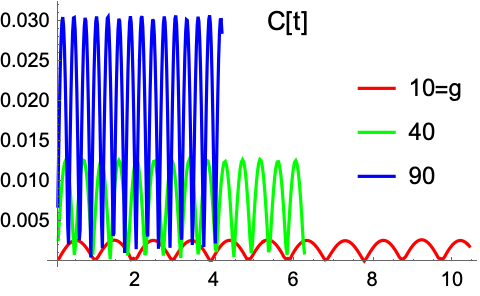}
\includegraphics[totalheight=25mm]{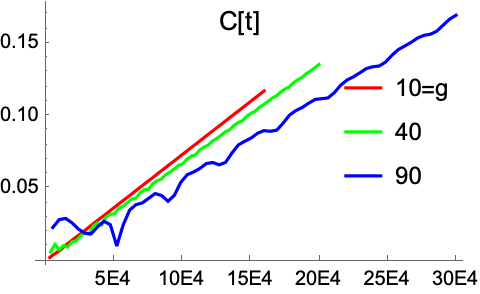}
\caption{Circuit complexity of the double rod pendulum. We set $\ell_1=\ell_2=1,m_1=m_2=1, g=10$, $40$ and $90$ respectively.}
\label{figCC}
\end{figure}
Our results are displayed in FIG.\ref{figCC}. From the figure, we see that in the short term evolution, the variation trend of CC exhibits manifestly periodical behaviour. However, in the long term limit, it grows linearly rather explicitly
\beq{}
\mathcal{C}(t)\rightarrow\lambda\cdot{}t
\eeq
This seems to provide a support evidence for the diagnosis of ref.\cite{Qu2021ius,Ali2019zcj} of quantum chaos which says that, the instability of classic system such as inverted harmonic oscillator on initial perturbations will cause linear growth of CC when the system is quantized. However, we must note that the parameter $\lambda$ estimated this way are inversely proportional to the strength of $g$ or $\hbar^{-2}$, in contrasts with the Lyapnov parameter estimated from FIG.\ref{figLypExpAsFuncOfg}. At the same time, we also see similar early-oscillation later linear-growth of CC defined through \eqref{aliCCdefinition}-\eqref{aliGRGTdefinition} even for exactly integrable system such as single pendulum. This implies that either the CC defined through \eqref{evolUDRP}-\eqref{xiDRP} and \eqref{aliCCdefinition}-\eqref{aliGRGTdefinition} is not a good diagnosis for quantum chaos or the double rod pendulum is not a strong chaotic system even at classic levels.

\section{Conclusion}
\label{secConcl}

In this work, we investigated the canonical quantisation of the standard double rod pendulum with the aim of developing a new lab to evaluate or test various diagnosis for quantum chaos. We first point out that in a quantised double rod pendulum, the strength of external field driving the pendulum and the planck constant always appear as the combination $g\hbar^{-2}$, so that the effect of enhancing $g$ is equivalent to that of weakening $\hbar$ or the quantum effects and vice versa. This means that the double rod pendulum is indeed an ideal laboratory to explore the way how classic chaos changes at quantum level, of course in isolated environment. We observed that as one weakens $g$ or strengthens $\hbar$, the scrambling time of the system increases, so its chaos feature weakens correspondingly.

We then solved the eigenstate Schr\"dinger equation of the system with spectral analysis method and get $10^4$ lowest lying eigenvalues and eigenstate wave functions with relative precision to at least $10^{-4}$. From these eigenvalues, we observed that the very popular NNSD diagnosis for quantum chaos fails in the double rod pendulum. That is, the Nearest Neighboring eigenvalue Spacing's Distribution of quantum double rod pendulum follows not that of Gaussian Orthogonal matrix Ensemble. Instead it obeys an explicitly poisson distribution. This implies that either NNSD is not a necessary condition for a quantum system to exhibit chaotic feature at classic limit or the double rod pendulum is not a strong chaotic system at classic levels. Nevertheless, we observed that the Next Nearest Neighboring eigenvalue Spacing Distribution (NNNSD) of the quantum double rod pendulum has GOE feature. 

With the eigenvalue and eigenstate wave functions obtained through the spectral analysis method, we calculated the OTOC F-function $\langle{}q_1(t)p_1(0)q_1(t)p_1(0)\rangle$ and C-function $\langle[q_1(t),p_1(0)]^2\rangle$ of the system numerically. We observed that the OTOC in the double rod pendulum exhibit manifestly early time exponential growth and late time constance approaching behaviour. We also observed that the feature of early time growth is consistent with the classic definition of chaos and the quantum version Lyapnov exponent is well bellow the upper bound of MSS conjecture. So all our calculations related with the OTOC diagnosis is consistent with the conclusion quotient in the literature.

We then generalised the definition of and calculated numerically the circuit complexity of reference \cite{Ali2019zcj}. Our results indicate that the CC we defined exhibits almost exactly periodic behaviour during the short time evolution but linearly growing feature in the long time limit. Similar with the conclusion concerning the NNSD diagnosis, the CC diagnosis also implies that, either the double rod pendulum is not a strong chaotic system at the classic level, or the linear growth of CC in the long term evolution limit are not a good diagnosis for quantum chaos.

\section*{Acknowledgements}
This work is supported by NSFC grant no. 11875082 \url{https://search.crossref.org/funding}, NSFS grant No. 2022NSFSC1806 and Fundamental Research Founds of China West Normal University (No.22kA005). %\href{https://search.crossref.org/funding}.
We thank very much for the warm hosts provided by Vitor Cardoso at Niels Bhor Institute and the support from Villum Investigator program supported by the VILLUM Foundation (grant no. VIL37766) and the DNRF Chair program (grant no. DNRF162) by the Danish National Research Foundation.

\end{document}